\documentclass[11pt,draftcls]{IEEEtran}
\onecolumn

\usepackage[compress]{cite}

\usepackage[T1]{fontenc}
\usepackage[mathcal]{euscript}
\usepackage{amsmath,amstext,amsfonts,amssymb,bm}

\usepackage[dvips]{graphicx}
\DeclareGraphicsExtensions{.eps}

\usepackage{stfloats}
\usepackage{hyperref}

\def \A {\mathbf{A}} \def \D {\mathbf{D}}  \def \H {\mathbf{H}} \def \Hb {\tilde{\H}}  \def \I {\mathbf{I}}   \def \n {\mathbf{n}} \def \S {\mathbf{S}}   \def \U {\mathbf{U}} \def \V {\mathbf{V}} \def \x {\mathbf{x}} \def \x {\mathbf{x}} \def \y {\mathbf{y}}
\def \PP {\mathbb{P}}
\def \Tr {\mathrm{Tr}}
\def \eps {\varepsilon}
\def \mhs {\hspace{-2pt}} \def \bhs {\hspace{-8pt}} \def \fhs {\hspace{7pt}}

\newcommand{\bs}{\boldsymbol}
\newtheorem{theorem}{Theorem}
\begin{document}

\title{Diversity of the MMSE receiver in flat fading and frequency selective MIMO channels at fixed rate}

\author{
Florian Dupuy\IEEEauthorrefmark{1}\IEEEauthorrefmark{2} and Philippe Loubaton\IEEEauthorrefmark{2}
\medskip \\
\IEEEauthorrefmark{1} Thales Communication EDS/SPM, 92704 Colombes (France) \\
\IEEEauthorrefmark{2} Universit\'e Paris Est, IGM LabInfo, UMR-CNRS 8049, 77454 Marne-la-Vall\'ee (France),
\medskip \\
Telephone: +33 146 132 109, Fax: +33 146 132 555, Email: fdupuy@univ-mlv.fr \\
Telephone: +33 160 957 293, Fax: +33 160 957 755, Email: loubaton@univ-mlv.fr
}

\maketitle

\begin{abstract}
  In this contribution, the evaluation of the diversity of the MIMO MMSE receiver is addressed for finite rates in both flat fading channels and frequency 
  selective fading channels with cyclic prefix. It has been observed recently that in contrast with the other MIMO receivers, the MMSE receiver has a diversity depending 
  on the aimed finite rate, and that for sufficiently low rates the MMSE receiver reaches the full diversity - that is, the diversity of the ML receiver.
  This behavior has so far only been partially explained. The purpose of this paper is to provide complete proofs for flat fading MIMO channels, and to 
  improve the partial existing results in frequency selective MIMO channels with cyclic prefix. 
\end{abstract}


\IEEEpeerreviewmaketitle

\section{Introduction}
The diversity-multiplexing trade-off (DMT) introduced by \cite{zheng2003diversity} studies the diversity function of the multiplexing gain in the high SNR regime.
\cite{kumar2009asymptotic} showed that the MMSE linear receivers, widely used for their simplicity, exhibit a largely suboptimal DMT in flat fading MIMO channels.
Nonetheless, for a finite data rate (i.e. when the rate does not increase with the signal to noise ratio), the MMSE receivers take several diversity values, depending on the aimed rate, as noticed earlier in \cite{hedayat2005linear}, and also in \cite{hedayat2004outage,tajer2007diversity} for frequency-selective MIMO channels.
In particular they achieve full diversity for sufficiently low data rates, hence their great interest.
This behavior was partially explained in \cite{kumar2009asymptotic,mehana2010diversity} for flat fading MIMO channels and in \cite{mehana2011diversity} for frequency-selective MIMO channels.
Indeed the proof of the upper bound on the diversity order for the flat fading case given in \cite{mehana2010diversity} contains a gap, and 
the approach of \cite{mehana2010diversity} based on the Specht bound seems to be unsuccessfull.
As for MIMO frequency selective channels with cyclic prefix, \cite{mehana2011diversity} only derives the diversity in the particular case of a number of channel taps equal to the transmission data block length, and claims that this value provides an upper bound in more realistic cases, whose expression is however not explicitly given.
In this paper we provide a rigorous proof of the diversity for MMSE receivers in flat fading MIMO channels for finite data rates.
We also derive the diversity in MIMO frequency selective channels with cyclic prefix for finite data rates if the transmission data block length is large enough. 
Simulations corroborate our derived diversity in the frequency selective channels case.

\section{Problem statement}
\begin{figure*}[t]
 \centering
 \includegraphics[width=6.5in]{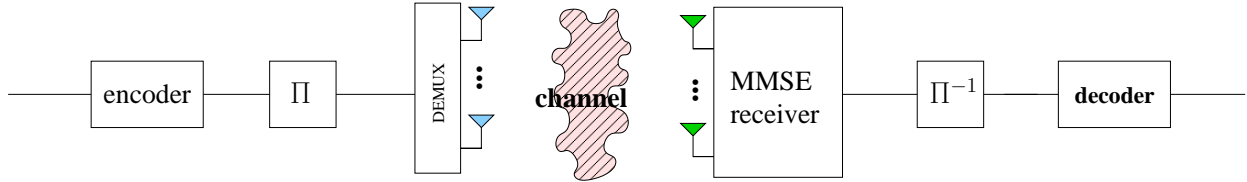}
 \caption{Considered MIMO system}
 \label{fig:scheme}
\end{figure*}

We consider a MIMO system with $M$ transmitting, $N \geq M$ receiving antennas,
with coding and ideal interleaving at the transmitter, and with
a MMSE linear equalizer at the receiver, followed by a de-interleaver and a decoder (see Fig. \ref{fig:scheme}).
We evaluate in the following sections the achieved diversity by studying the outage probability, that is the probability that the capacity does not support the target data rate, at high SNR regimes.
We denote $\rho$ the SNR, $I$ the capacity and $R$ the target data rate.
We use the notation $\doteq$ for {\it exponential equality} \cite{zheng2003diversity}, i.e.
\begin{equation}
 f(\rho) \doteq \rho^d
 \Leftrightarrow
 \lim_{\rho \to \infty} \frac{\log f(\rho)}{\log \rho}= d,
 \label{eq:exp-equ}
\end{equation}
and the notations $\dot\leq$ and $\dot\geq$ for exponential inequalities, which are similarly defined.
We note $\log$ the logarithm to base $2$.

\section{Flat fading MIMO channels}
In this section we consider a flat fading MIMO channel.
The output of the MIMO channel is given by
\begin{equation}
  \y= \sqrt{\frac{\rho}{M}} \H \x + \n,
\end{equation}
where $\n \sim \mathcal{CN}(\bs{0},\I_N)$ is the additive white Gaussian noise and
$\x$ the channel input vector,
$\H$ the $N \times M$ channel matrix with i.i.d. entries $\sim \mathcal{CN}(0,1)$.
\medskip
\begin{theorem}
  For a rate $R$ such that
  $\log \frac{M}{m} < \frac{R}{M} < \log \frac{M}{m-1}$,
  with $m \in \{ 1, \ldots, M \}$,
  the outage probability verifies
  \begin{equation}
  \PP(I<R) \doteq \rho^{-m(N-M+m)},
  \end{equation}
  that is, a diversity of $m(N-M+m)$.  
\end{theorem}
\medskip

Note that for a rate $R < M \log \frac{M}{M-1}$ (i.e. $m=M$) full diversity $MN$ is attained, while for a rate $R > M \log M$ the diversity corresponds to the one derived by DMT approach.
This result was stated by \cite{mehana2010diversity}.
Nevertheless the proof of the outage lower bound in \cite{mehana2010diversity} omits that the event noted $\mathcal{B}_a$ is not independent from the eigenvalues of $\H^H\H$, hence questioning the validity of the given proof.
We thus provide an alternative proof based on an approach suggested by the analysis of \cite{kumar2009asymptotic} in the case 
where $R = r \log \rho$ with $r  > 0$. 

\medskip
\begin{proof}
 The capacity $I$ of the MIMO MMSE considered system is given by
 \[
   I = \sum_{j=1}^M \log ( 1 + \beta_j),
 \]
 where $\beta_j$ is the SINR for the $j$th stream:
 \[
   \beta_j= \frac{1}{\left(  \left[ \I + \frac{\rho}{M} \H^*\H  \right]^{-1} \right)_{jj} } - 1.
 \]
 We lower bound in the first place $\PP(I<R)$ and prove in the second place that the bound is tight by upper bounding $\PP(I<R)$ with the same bound.
 
 \subsection{Lower bound of the outage probability}
 \label{sec:lowB_flat}
 We here assume that $R/M>\log (M/m$).
 In order to lower bound $\PP(I<R)$ we need to upper bound the capacity $I$. Using Jensen's inequality on function $x \mapsto \log x$ yields
 \begin{align}
    I
    &\leq M \log  \Bigg[ \frac{1}{M} \sum_{j=1}^M  \left( 1+\beta_j \right) \Bigg] \label{ineq:jensen1}
    \\
    &= M \log  \Bigg[ \frac{1}{M} \sum_{j=1}^M   \bigg( \left[  \left( \I + \frac{\rho}{M} \H^*\H  \right)^{-1} \right]_{jj} \bigg)^{-1} \Bigg]. \label{ineq:logconcave}
 \end{align}
 We note $\H^*\H= \U^*\Lambda\U$ the SVD of $\H^*\H$ with $\Lambda=\mathrm{diag}(\lambda_1,\ldots,\lambda_M)$, $\lambda_1 \leq \lambda_2 \ldots \leq \lambda_M$.
 We recall that the $(\lambda_k)_{k=1, \ldots, M}$ are independent from the entries of matrix $\U$ and that $\U$ is a Haar distributed unitary random matrix, i.e. the probability distribution of $\U$ is invariant by left (or right) multiplication by deterministic matrices.
 Using this SVD we can write
 \begin{equation}
  \frac{1}{M} \sum_{j=1}^M    \bigg( \left[  \left( \I + \frac{\rho}{M} \H^*\H  \right)^{-1} \right]_{jj} \bigg)^{-1} 
   = \frac{1}{M} \sum_{j=1}^M   \bigg(  \sum_{k=1}^M \frac{|\U_{kj}|^2}{1+ \frac{\rho}{M} \lambda_k }  \bigg)^{-1}. \label{eq:sumSINR}
 \end{equation}
 
 \subsubsection{\texorpdfstring{Case $m=1$}{Case m=1}}
 In order to better understand the outage probability behavior, we first consider the case $m=1$.
 In this case $R/M>\log M$.
 We review the approach of \cite[III]{kumar2009asymptotic},
 which consists in upper bounding \eqref{eq:sumSINR} by
 $\left( 1+ \frac{\rho}{M} \lambda_1 \right) \frac{1}{M} \sum_{j=1}^M  \frac{1}{|\U_{1j}|^2}$, 
 as
 $\sum_{k=1}^M \frac{|\U_{kj}|^2}{1+ \frac{\rho}{M} \lambda_k } \geq \frac{|\U_{1j}|^2}{1+\frac{\rho}{M}\lambda_1}$.
 Using this bound in \eqref{ineq:logconcave} gives
 \[
   I \leq M \log  \bigg[  \left( 1 + \frac{\rho}{M} \lambda_1 \right) \frac{1}{M} \sum_{j=1}^M \frac{1}{|\U_{1j}|^2} \bigg].
 \]
 Therefore
 \[
   \Big( \left( 1 + \frac{\rho}{M} \lambda_1 \right) \frac{1}{M} \sum_{j=1}^M \frac{1}{|\U_{1j}|^2} < 2^{R/M}  \Big) \subset (I<R).
 \]
 In order to lower bound $\PP(I<R)$, \cite{kumar2009asymptotic} introduced the set 
 \[
    \mathcal{A}_1 =  \bigg\{ \frac{1}{M} \sum_{j=1}^M \frac{1}{|\U_{1j}|^2} < M + \eps \bigg\}
 \]
 for $\eps > 0$.
 Then,
 \begin{align*}
  \PP(I <R)
  & \geq \PP\left( (I<R) \cap \mathcal{A}_1 \right)
  \\
  & \geq \PP \bigg[   \bigg( \left( 1 + \frac{\rho}{M} \lambda_1 \right) \frac{1}{M} \sum_{j=1}^M \frac{1}{|\U_{1j}|^2} < 2^{R/M}  \bigg) \cap \mathcal{A}_1\bigg]
  \\
  & \geq \PP \left[   \left( 1 + \frac{\rho}{M} \lambda_1 < \frac{2^{R/M}}{M+\eps} \right) \cap \mathcal{A}_1\right]
  \\
  & =  \PP(\mathcal{A}_1) \cdot \PP\left[  1 + \frac{\rho}{M} \lambda_1  < \frac{2^{R/M}}{M+\eps} \right],
 \end{align*}
 where the last equality comes from the independence between eigenvectors and eigenvalues of Gaussian matrix $\H^*\H$.
 It is shown in \cite[Appendix A]{kumar2009asymptotic} that
 $\PP(\mathcal{A}_1) \neq 0$.
 Besides, as we supposed
 $2^{R/M}>M$,
 we can take $\eps$ such that
 $\frac{2^{R/M}}{M+\eps}>1$, ensuring that $\PP\Big[ \left( 1 + \frac{\rho}{M} \lambda_1 \right) < \frac{2^{R/M}}{M+\eps} \Big] \neq 0$.
 Hence there exists $\kappa>0$ such that
 \[
   \PP(I<R)  \ \dot\geq  \ \PP\left( \lambda_1 < \frac{\kappa}{\rho} \right),
 \]
 which is asymptotically equivalent to $\rho^{-(N-M+1)}$ in the sense of \eqref{eq:exp-equ} (see, e.g., \cite[Th. II.3]{jiang2011performance}).

 \subsubsection{\texorpdfstring{General case $1 \leq m\leq M$}{General case 1<=m<= M}}
 By the same token as for $m=1$ we now consider the general case --
 we recall that we assumed that $\log (M/m) < R/M$.
 We first lower bound $\sum_k\frac{|\U_{kj}|^2}{1+ \frac{\rho}{M} \lambda_k}$ which appears in \eqref{eq:sumSINR} by the $m$ first terms of the sum and then use Jensen's inequality applied on $x \mapsto x^{-1}$, yielding
 \begin{align*}
  \sum_{k=1}^M \frac{|\U_{kj}|^2}{1+ \frac{\rho}{M} \lambda_k}
  &\geq \sum_{k=1}^m \frac{|\U_{kj}|^2}{1+ \frac{\rho}{M} \lambda_k}
  \\
  &\geq \frac{\left( \sum_{l=1}^m |\U_{lj}|^2 \right)^2}{\sum_{k=1}^m |\U_{kj}|^2 \left( 1+ \frac{\rho}{M} \lambda_k\right)}.
 \end{align*}
 Using this inequality in \eqref{eq:sumSINR}, we obtain that
 \begin{align}
    \frac{1}{M} \sum_{j=1}^M    \bigg( \left[  \left( \I + \frac{\rho}{M} \H^*\H  \right)^{-1} \right]_{jj} \bigg)^{-1}
     \leq \frac{1}{M} & \sum_{j=1}^M \frac{\sum_{k=1}^m |\U_{kj}|^2\left( 1+ \frac{\rho}{M} \lambda_k\right)}{\left( \sum_{l=1}^m |\U_{lj}|^2 \right)^2} \notag
     \\
     & = \sum_{k=1}^m  \left( 1+ \frac{\rho}{M} \lambda_k\right)  \delta_k(\U), \label{ineq:intr_delta}
 \end{align}
 where
 $\delta_k(\U) = \frac{1}{M} \sum_{j=1}^M  \frac{|\U_{kj}|^2}{\left( \sum_{l=1}^m |\U_{lj}|^2 \right)^2}$.
 Equation \eqref{ineq:intr_delta}, together with \eqref{ineq:logconcave}, yields the following inclusion:
 \begin{align*}
  \Bigg( \sum_{k=1}^m  \delta_k(\U) \left( 1+ \frac{\rho}{M} \lambda_k\right)  <   2^{R/M} \Bigg)
  \subset (I<R).
 \end{align*}
 Similarly to the case $m=1$, we introduce the set $\mathcal{A}_m$ defined by
 \[
    \mathcal{A}_m =  \left\{ \delta_k(\U) < \frac{M}{m^2} + \eps, \ 
    k=1,\ldots,m \right\}
 \]
 for $\eps > 0$.
 We now use this set to lower bound $\PP(I<R)$.
 \begin{align*}
  \PP(I<R) & \geq \PP\left( (I<R) \cap \mathcal{A}_m \right)
  \\
  & \geq \PP \bigg[   \Bigg( \sum_{k=1}^m  \delta_k(\U) \left( 1+ \frac{\rho}{M} \lambda_k\right)  <   2^{R/M} \Bigg) \cap \mathcal{A}_m\bigg]
  \\
  & \geq \PP \left[   \Bigg( \sum_{k=1}^m \left( 1+ \frac{\rho}{M} \lambda_k\right)  <  \frac{2^{R/M}}{\frac{M}{m^2} + \eps} \Bigg) \cap \mathcal{A}_m\right]
  \\
  &= \PP(\mathcal{A}_m) \cdot \PP\left[  \sum_{k=1}^m \left( 1 + \frac{\rho}{M} \lambda_k  \right) < \frac{2^{R/M}}{\frac{M}{m^2}+\eps} \right].
 \end{align*}
  The independence between eigenvectors and eigenvalues of Gaussian matrix $\H^*\H$ justifies the last equality.
  As we assumed that
  $\log(M/m) < R/M$, that is $m < \frac{2^{R/M}}{M/m^2}$,
  we can choose $\eps$ such that
  $m < \frac{2^{R/M}}{M/m^2 +\eps}$.
  That ensures that $\PP\left[  \sum_{k=1}^m \left( 1 + \frac{\rho}{M} \lambda_k  \right) < \frac{2^{R/M}}{ M/m^2 +\eps} \right] \neq 0$.
  We show in Appendix \ref{apx:prob_sum_first_ev} that this probability is asymptotically equivalent to $\rho^{-m(N-M+m)}$ in the sense of \eqref{eq:exp-equ}, leading to
  \begin{equation}
    \PP(I<R) \ \dot\geq \  \frac{\PP(\mathcal{A}_m)}{\rho^{m(N-M+m)}}.
    \label{ineq:P_I<R}
  \end{equation}

  We still need to prove that $\PP(\mathcal{A}_m) \neq 0$. 
  Any Haar distributed random unitary matrix can be parameterized by $M^2$ independent angular random variables
  $(\alpha_1, \ldots, \alpha_{M^2})=\bs\alpha$ whose probability distributions are almost surely positive (see \cite{dita2003factorization, lundberg2004haar} and Appendix \ref{apx:ang_par}).
  We note $\Phi_m$ the functions such that $\U=\Phi_m(\bs\alpha)$.
  Consider a deterministic unitary matrix $\U_*$ such that $|(\U_*)_{ij}|^2 = \frac{1}{M} \ \forall i,j$,
  and denote by $\bs\alpha_*$ a corresponding $M^2$ dimensional vector.
  It is straightforward to check that $\delta_k \circ \Phi_m (\alpha_*) = M / m^2$.
  Functions $\bs\alpha \mapsto (\delta_k \circ \Phi_m)(\bs\alpha)$ are continuous at point $\bs\alpha_*$ for $1 \leq k \leq m$ and therefore there exists $\eta>0$ such that the ball
  $\mathcal{B} \left( \bs\alpha_*, \eta \right)$ is included in the set $\left\{ \bs\alpha, \  (\delta_k \circ \Phi_m)(\alpha) < \frac{M}{m^2}+\eps, \ k=1, \ldots, m \right\}$.
  We have therefore $\PP(\mathcal{A}_m) \neq 0$ as
  \begin{align*}
   \PP(\mathcal{A}_m)
   & = \int_{\left\{ (\delta_k \,\circ\, \Phi_m) (\bs\alpha) < \frac{M}{m^2} + \eps, \, k=1, \ldots, m \right\}} p(\bs\alpha) d\bs\alpha
   \\
   & > \int_{\mathcal{B} \left( \bs\alpha_*, \eta \right)} p(\bs\alpha) d\bs\alpha > 0
  \end{align*}
  Coming back to \eqref{ineq:P_I<R}, we eventually have
  \[
   \PP(I<R) \ \dot\geq \ \frac{1}{\rho^{m(N-M+m)}},
  \]
  that is the diversity of the MMSE receiver is upper bounded by $m(N-M+m)$.
  
  \subsection{Upper bound of the outage probability}
  \label{sec:upper-flat}
  We now conclude by studying the upper bound of the outage probability, showing that $m(N-M+m)$ is also a lower bound for the diversity.
  Note that this lower bound has been derived in \cite{kumar2009asymptotic, mehana2010diversity} using however rather informal arguments; we provide a more rigorous proof here for the sake of completeness.
  
  We now assume that
  $R/M < \log (M/(m-1))$, i.e. $m-1 < M 2^{-R/M}$.
  Using Jensen inequality on function $y \mapsto \log(1/y)$, the capacity $I$ can be lower bounded:
  \begin{align*}
    I
    &= - \sum_{j=1}^M \log \left( \left[ \left( \I + \frac{\rho}{M} \H^*\H  \right)^{-1} \right]_{jj} \right)
    \\
    &\geq -M \log \left( \frac{1}{M} \Tr \left[ \left( \I + \frac{\rho}{M} \H^*\H  \right)^{-1} \right] \right),
  \end{align*}
  which leads to an upper bound for the outage probability:
  \begin{equation}
   \PP(I<R) 
   \leq \PP \left[ \Tr \left[ \left( \I + \frac{\rho}{M} \H^*\H  \right)^{-1} \right] > M 2^{-R/M} \right].
   \label{ineq:PP-B0}
  \end{equation}
  We need to derive the probability in the right-hand side of the above inequality.
  Noting
  $\mathcal{B}_0= \left\{ \lambda_1 \leq \lambda_2 \ldots \leq \lambda_M, \ \sum_{k=1}^M \left( 1 +  \frac{\rho}{M} \lambda_k  \right)^{-1} > M 2^{-R/M} \right\}$,
  \begin{align}
    \PP  \bigg[ \Tr \Big[ \Big( \I + & \frac{\rho}{M} \H^*\H  \Big)^{-1} \Big] > M 2^{-R/M} \bigg] \notag
    \\
    & = \int_{\mathcal{B}_0} p(\lambda_1, \ldots, \lambda_M) d\lambda_1 \ldots d\lambda_M.
    \label{eq:int_B0_PP}
  \end{align}
  
  We now introduce $\mu_m =  \sup_{(\lambda_1, \ldots, \lambda_M) \in \mathcal{B}_0}\{ \rho \,\lambda_m \}$
  and prove by contradiction that $\mu_m < +\infty$.
  If $\mu_m = +\infty$, there exists a sequence $(\lambda_1^{(n)}, \lambda_2^{(n)}, \ldots, \lambda_M^{(n)})_{n\in\mathbb{N}}$ such that
  $\lambda_k^{(n)} \rightarrow +\infty$ for any $k \geq m$.
  Besides,
  \[
   M 2^{-R/M} \mhs < \sum_{k=1}^M \mhs \Big(  1+\frac{\rho}{M}\lambda^{(n)}_k \Big)^{-1} \mhs \leq (m-1) + \sum_{k=m}^M \mhs \Big(1+\frac{\rho}{M}\lambda^{(n)}_k\Big)^{-1}.
  \]
  In particular $M 2^{-R/M} < (m-1) + \sum_{k=m}^M \mhs \big(1+\frac{\rho}{M}\lambda^{(n)}_k\big)^{-1} $,
  which, taking the limit when $n \rightarrow +\infty$, leads to $m-1 \geq M 2^{-R/M}$, a contradiction with the assumption
  $m-1 < M 2^{-R/M}$. Hence, $\mu_m < +\infty$.
  
  We introduce the set $\mathcal{B}_1= \{ \lambda_1 \leq \lambda_2 \ldots \leq \lambda_M, \, 0 < \lambda_k \leq \frac{\mu_m}{\rho}, \, k=1,\ldots,m \}$,
  which verifies $\mathcal{B}_0 \subset \mathcal{B}_1$.
  Using \eqref{ineq:PP-B0} and \eqref{eq:int_B0_PP}, this implies that
  \[
    \PP(I<R) \leq \int_{\mathcal{B}_1} p(\lambda_1, \ldots, \lambda_M) d\lambda_1 \ldots d\lambda_M,
  \]
  which is shown to be asymptotically smaller than
  $\rho^{-m(N-M+m)}$ in the sense of \eqref{eq:exp-equ}
  in Appendix \ref{apx:prob_first_ev_bounded}.
  The diversity is thus lower bounded by $m(N-M+m)$, ending the proof.
\end{proof}

\section{Frequency selective MIMO channels with cyclic prefix}
\label{sec:freqsel}
We consider a frequency selective MIMO channel with $L$ independent taps.
We consider a block transmission cyclic prefix scheme, with a block length of $K$.
The output of the MIMO channel at time $t$ is given by
\begin{align*}
 \y_t =\sqrt{\frac{\rho}{ML}}\,\sum_{l=0}^{L-1} \H_l \x_{t-l} +\n_t
 = \sqrt{\frac{\rho}{ML}}\,[\H(z)] \x_t +\n_t
\end{align*}
where
$\x_t$ is the channel input vector at time $t$,
$\n_t \sim \mathcal{CN}(\bs{0}, \I_N)$ the additive white Gaussian noise,
$\H_l$ is the $N \times M$ channel matrix associated to $l^{\mathrm{th}}$ channel tap, for $l \in \{0,\ldots, L-1\}$,
and $\H(z)$ denotes the transfer function of the discrete-time equivalent channel defined by
\[
  \H(z) = \sum_{l=0}^{L-1} \H_l \, z^{-l}.
\]
We make the common assumption that the entries of $\H_l$ are i.i.d and $\mathcal{CN}(0,1)$ distributed.
We can now state the second diversity theorem of the paper.
\medskip
\begin{theorem}
Assume that the non restrictive condition 
$K > {M^{2}(L-1)}$
holds, ensuring that $\log \frac{M}{m} < -\log\big(\frac{m-1}{M}+\frac{(L-1)(M-(m-1))}{K}\big)$ for any $m=1, \ldots, M$.
Then, for a rate $R$ verifying
\begin{equation}
 \hspace{-11pt}
 \textstyle
  \log \frac{M}{m}
  < \frac{R}{M} <
  -\log \left(\frac{m-1}{M} + \frac{(L-1)(M-(m-1))}{K}  \right),
 \label{eq:bounds_R_fsel}
\end{equation}
$m \in \{ 1, \ldots, M \}$, the outage probability verifies 
\begin{equation}
 \PP(I<R) \doteq \rho^{-m(LN-M+m)},
\end{equation}
that is a diversity of $m(LN-M+m)$.
\end{theorem}
\medskip

The diversity of the MMSE receiver is thus $m(LN-M+m)$, corresponding to a flat fading MIMO channel with $M$ transmit antennas and 
$LN$ receive antennas. For a large block length $K$, the upper bound for rate $R$ is close to the bound of the previous flat fading case $\log \frac{M}{m-1}$.
Concerning data rates verifying
$
 -\log\big(\frac{m-1}{M} + \frac{L-1}{K} (M-(m-1))\big)
 < \frac{R}{M} <
 \log \frac{M}{m-1},
$
the $m(LN-M+m)$ diversity is only an upper bound;
nevertheless the diversity is also lower bounded by $(m-1)(LN-M+(m-1))$.

\medskip
\begin{proof}
 Similarly to previous section the capacity of the MIMO MMSE system is written
  $
   I = \sum_{j=1}^M \log ( 1 + \beta_j),
  $
 where $\beta_j$ is the SINR for the $j$th stream of $\x_t$.
 It is standard material that in MIMO frequency selective channel with cyclic prefix the SINR of the MMSE receiver is given by
 \begin{equation}
  \beta_j = \frac{1}{ \frac{1}{K}\sum_{k=1}^K \left[ \left( \S\left(\frac{k-1}{K}\right)  \right)^{-1} \right]_{jj} } -1,
  \label{eq:SINR_freqsel}
 \end{equation}
 where $\S(\nu)=  \I_N + \frac{\rho}{M} \H(e^{2 i \pi \nu})^*\H(e^{2 i \pi \nu}) $.
 
 \subsection{Lower bound for the outage probability}
 We assume that $R/M > \log(M/m)$.
 
 One can show that function $\A \mapsto (\A^{-1})_{jj}$, defined over the set of positive-definite matrices, is convex.
 Using Jensen's inequality then yields
 \begin{align*}
   \frac{1}{K}\sum_{k=1}^K {\textstyle \left[ \left( \S\left(\frac{k-1}{K}\right)  \right)^{-1} \right]_{jj} }
   & \geq \bigg( \bigg[ \frac{1}{K} \sum_{k=1}^K {\textstyle  \S\left(\frac{k-1}{K}\right) }\bigg]^{-1} \bigg)_{jj}
   \\
   & = \bigg( \bigg[ \I_N +  \sum_{l=0}^{L-1} \frac{\rho}{M} \H_l^*\H_l \bigg]^{-1} \bigg)_{jj}.
 \end{align*}
 The last equality follows from the fact that $\frac{1}{K}\sum_{k=1}^K e^{2 i\pi \frac{k-1}{K}(l-n)}=\delta_{ln}$.
 Using this inequality in the SINR expression \eqref{eq:SINR_freqsel} gives
 \[
  1+\beta_j \leq \bigg( \bigg( \bigg[ \I_N +  \sum_{l=0}^{L-1} \frac{\rho}{M} \H_l^*\H_l  \bigg]^{-1} \bigg)_{jj} \bigg)^{-1}.
 \]
 We now come back to the capacity $I$ of the system;
 similarly to \eqref{ineq:jensen1}, using Jensen's inequality yields
 \begin{align*}
  I 
  & \leq M \log  \Bigg[ \frac{1}{M} \sum_{j=1}^M  \left( 1+\beta_j \right) \Bigg]
  \\
  & \leq M \log \Bigg[  \frac{1}{M} \sum_{j=1}^M \bigg( \bigg( \bigg[ \I_N +  \frac{\rho}{M} \sum_{l=0}^{L-1} \H_l^*\H_l  \bigg]^{-1} \bigg)_{jj} \bigg)^{-1} \,\Bigg].
 \end{align*}
 We can now use the results of section \ref{sec:lowB_flat} by simply replacing $N \times M$ matrix $\H$ in \eqref{ineq:logconcave} by $LN \times M$ matrix $\Hb=[\H_0^T, \H_1^T, \ldots, \H_{L-1}^T]^T$.
 They lead to the following lower bound for the outage capacity, for a rate $R$ verifying $R/M > \log (M/m)$:
 \[
   \PP(I<R) \ \dot\geq \ \frac{1}{\rho^{m(LN-M+m)}}.
 \]

 \subsection{Upper bound for the outage probability}
  We assume that $\frac{R}{M} < -\log\big(\frac{m-1}{M}+\frac{(L-1)(M-(m-1))}{K}\big)$, that is
  $2^{-R/M} < \frac{m-1}{M} \mhs + \mhs \frac{L-1}{K} (M \mhs - \mhs (m-1))$.
  
  We first derive a lower bound for the capacity $I$.
 \begin{align*}
  I & = - \sum_{j=1}^M \log \left( \frac{1}{K}\sum_{k=1}^K \left( \left[ {\textstyle \S\left(\frac{k-1}{K}\right)} \right]^{-1} \right)_{jj}  \right)
  \\
  & \geq - M  \log \left( \frac{1}{KM} \sum_{k=1}^K \Tr\left( \left[ {\textstyle \S\left(\frac{k-1}{K}\right)} \right]^{-1} \right) \right)
 \end{align*}
 The latter inequality follows once again from Jensen's inequality on function $x \mapsto \log x$.
 
 We now analyze $\Tr\left(\S(\nu)^{-1}\right)$.
 To that end, we write $LN \times M$ matrix $\Hb=[\H_0^T, \ldots, \H_{L-1}^T]^T$ under the form
 $\Hb=\bs\Theta (\Hb^*\Hb)^{1/2}$,
 where $\bs\Theta=[\bs\Theta_0^T, \ldots, \bs\Theta_{L-1}^T]^T$ and $\bs\Theta^*\bs\Theta=\I_M$.
 Besides, we note $\U^*\Lambda\U$ the SVD of $\Hb^*\Hb$ with
 $\Lambda=\mathrm{diag}(\lambda_1,\ldots,\lambda_M)$,
 $\lambda_1 \leq \ldots \leq \lambda_M$.
 Hence,
 \[
  \H(e^{2 i\pi \nu})=\bs\Theta(e^{2 i\pi \nu})\U^*\Lambda^{1/2}\U,
 \]
 where
 $\bs\Theta(z)=\sum_{l=0}^{L-1}\bs\Theta_l z^{-l}$.
 Using this parametrization,
 \begin{align*}
  \Tr\left(\S(\nu)^{-1}\right)
   &= \Tr \left[ \left( \I + \frac{\rho}{M} \U \bs\Theta^*(e^{2 i \pi \nu})\bs\Theta(e^{2 i \pi \nu}) \U^* \Lambda \right)^{-1} \right]
   \\
   &\leq \Tr \left[ \left( \I + \frac{\rho}{M} \gamma(e^{2 i \pi \nu}) \Lambda \right)^{-1} \right],
 \end{align*}
 where $\gamma(\nu)=\lambda_{\mathrm{min}}(\bs\Theta^*(e^{2 i \pi \nu})\bs\Theta(e^{2 i \pi \nu}))$.
 Coming back to the outage probability,
 \begin{align}
  \PP(I<R) \leq & \PP \mhs \left[  \frac{1}{K} \sum_{k=0}^{K-1} \sum_{j=1}^M \mhs \left( \mhs 1 + \frac{\rho \lambda_j }{M} \gamma \mhs\left(\frac{k}{K}\right) \mhs \right)^{\mhs-1} \mhs > M2^{-R/M}\right] \notag
  \\ 
  &= \PP \mhs \left[  \Hb \in \mathcal{B}_0 \right], \label{ineq:PP-fs}
 \end{align}
 where $\mathcal{B}_0 = \big\{ \Hb, \frac{1}{K} \sum_{k=0}^{K-1} \sum_{j=1}^M \mhs \big( 1 + \frac{\rho \lambda_j }{M} \gamma \mhs\left(\frac{k}{K}\right)  \big)^{-1} \mhs > M2^{-R/M} \big\}$.
 
 We now prove by contradiction that
 $\mu_m < +\infty$, where $\mu_m =\sup_{\Hb \in \mathcal{B}_0} \{ \rho \lambda_m \}$.
 If $\mu_m = +\infty$ 
 there exists a sequence of matrices $\Hb^{(n)} \mhs \in \mhs \mathcal{B}_0$ such that $\rho \lambda_m^{(n)} \rightarrow +\infty$.
 Besides,
 \begin{align}
  M 2^{-\frac{R}{M}}
  & < \frac{1}{K} \sum_{k=0}^{K-1} \sum_{j=1}^M \bigg( 1 + \frac{\rho \lambda_j^{(n)} }{M} \gamma^{(n)} \mhs\left(\frac{k}{K}\right)  \bigg)^{\mhs-1} \notag
  \\
  & \leq (m-1) + \frac{1}{K} \sum_{k=0}^{K-1} \sum_{j=m}^M \bigg( 1 + \frac{\rho \lambda_j^{(n)} }{M} \gamma^{(n)} \mhs\left(\frac{k}{K}\right)  \bigg)^{\mhs-1}
  \label{ineq:mat-seq}
 \end{align}
 As $\bs\Theta^{(n)}$ belongs to a compact we can extract a subsequence $\bs\Theta^{(\psi(n))}$ which converges towards a matrix $\bs\Theta_\infty$.
 For this subsequence, inequality \eqref{ineq:mat-seq} becomes
 \begin{equation}
  M 2^{-\frac{R}{M}}
   \leq (m-1) + \frac{1}{K} \sum_{k=0}^{K-1} \sum_{j=m}^M \bigg( 1 + \frac{\rho \lambda_j^{(\psi(n))} }{M} \gamma^{(\psi(n))} \mhs\left(\frac{k}{K}\right)  \bigg)^{-1}.
   \label{ineq:sumsum}
 \end{equation}
 Let  $\gamma_\infty$ be the function defined by $\gamma_\infty(\nu)=\lambda_{\mathrm{min}}(\bs\Theta^*_\infty(e^{2 i \pi \nu})\bs\Theta_\infty(e^{2 i \pi \nu}))$ and $k_1, \ldots, k_p$ be the integers for which $\gamma_{\infty}(k_j/K) = 0$.
 Then $\det \bs\Theta_\infty(z) = \det \big( \sum_{l=0}^{L-1} \bs\Theta_{\infty,l} z^{-l} \big)=0$ for all $z \in \big\{ e^{2 i \pi k_j/K}, j=1,\ldots,p \big\}$.
 Nevertheless, polynomial $z \mapsto \sum_{l=0}^{L-1} \bs\Theta_{\infty,l} z^{-l}$ has a maximum degree of $M(L-1)$, therefore $p \leq M(L-1)$.
 Inequality \eqref{ineq:sumsum} then leads to
 \begin{equation}
  M 2^{-\frac{R}{M}}
   \leq (m-1) +  \frac{M(L-1)}{K} 
    +\frac{1}{K}  \sum_{k\notin\{k_1, \ldots, k_p\}} \sum_{j=m}^M \bigg( 1 + \frac{\rho \lambda_j^{(\psi(n))} }{M} \gamma^{(\psi(n))} \left( \frac{k}{K} \right)  \bigg)^{\mhs-1} \label{ineq:M2-RMp}
 \end{equation}
 Moreover, if $k \notin \{k_1, \ldots, k_p\}$,
 $\lambda_j^{(\psi(n))} \gamma^{(\psi(n))}(\frac{k}{K}) \rightarrow +\infty$ for $j \geq m$,
 as $\gamma^{(\psi(n))} \mhs\left(\frac{k}{K}\right) \rightarrow \gamma_\infty \mhs\left(\frac{k}{K}\right) \neq 0$ for $k \notin \left\{ k_1, \ldots, k_p \right\}$.
 Therefore taking the limit of \eqref{ineq:M2-RMp} when $n \rightarrow +\infty$ gives
 \[
  M 2^{-\frac{R}{M}}
   \leq (m-1) +  \frac{M(L-1)}{K},
 \]
 which is in contradiction with the original assumption
 $2^{-R/M} < \mhs \frac{m-1}{M} \mhs + \mhs \frac{L-1}{K} (M \mhs - \mhs (m-1))$.
 Hence $\mu_m < +\infty$, and
 $\mathcal{B}_0 \subset \mathcal{B}_1= \{ \Hb, \rho \lambda_m(\Hb^*\Hb) < \mu_m \}$.
 Using \eqref{ineq:PP-fs}, we thus have
 \[
  \PP(I<R) \leq \PP(\Hb \in \mathcal{B}_1),
 \]
 which, by Appendix \ref{apx:prob_first_ev_bounded}, is asymptotically smaller than
 $\rho^{-m(NL-M+m)}$ in the sense of \eqref{eq:exp-equ}, therefore ending the proof.
\end{proof}

\section{Numerical Results}
\begin{figure}[t]
 \centering
 \includegraphics[width=3in]{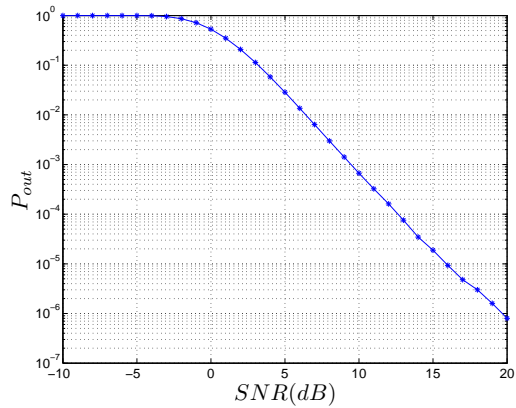}
 \caption{Outage probability of the MMSE receiver, L=2, K=64, M=N=2}
 \label{fig:Pout}
\end{figure}

We here illustrate the derived diversity in the frequency selective case.
In the conducted simulation we took a block length of $K=64$,
a number of transmitting and receiving antennas $M=N=2$,
$L=2$ channel taps
and an aimed data rate $R=3$~bits/s/Hz.
Rate $R$ then verifies \eqref{eq:bounds_R_fsel} with $m=1$, therefore the expected diversity is $LN-M+1=3$.
The outage probability is displayed on Fig. \ref{fig:Pout} as a function of SNR.
We observe a slope of $-10^{-3}$ per decade, hence a diversity of $3$, confirming the result stated in part \ref{sec:freqsel}.

\section{Conclusion}
  In this paper we provided rigorous proofs regarding the diversity of the MMSE receiver at fixed rate, in both flat fading and frequency selective MIMO channels.
  The higher the aimed rate the less diversity is achieved;
  in particular, for sufficiently low rates, the MMSE receiver achieves full diversity in both MIMO channel cases, hence its great interest.
  Nonetheless, in frequency selective channels, the diversity bounds are not tight for some specific rates; this could probably be improved.
  Simulations corroborated our results.

\appendices
\section{}
\label{apx:prob_sum_first_ev}
 We prove in this appendix that, for $b>0$, $\PP(\sum_{k=1}^m \rho \lambda_k < b) \ \dot\geq \ \rho^{-m(N-M+m)}$.
 
 We note $\mathcal{C}_m$ the set defined by
 $\mathcal{C}_m = \{ \lambda_1, \ldots, \lambda_m: \ 0 < \lambda_1 \leq \ldots \leq \lambda_m, \ \sum_{k=1}^m \rho \lambda_k < b \}$.
 As the $\lambda_i$ verify
 $0 < \lambda_1 \leq \ldots \leq \lambda_M$,
 we can write
 \begin{equation}
  \PP \Bigg( \sum_{k=1}^m \rho \lambda_k < b \Bigg)
  = \int_{(\lambda_1, \ldots, \lambda_m) \in \mathcal{C}_m} \int_{\lambda_m}^{+\infty} \bhs\ldots \int_{\lambda_{M-1}}^{+\infty} p_{M,N}(\lambda_1, \ldots, \lambda_M) \ d\lambda_1 \ldots d\lambda_M,
  \label{eq:prob_sum_first_ev}
 \end{equation}
 where
 $p_{M,N}: \mathbb{R}^M \rightarrow \mathbb{R}$
 is the joint probability density function of the ordered eigenvalues of a $M \times M$ Wishart matrix with scale matrix $\I_M$ and $N$ degrees of freedom, given by (see, e.g., \cite{zheng2003diversity}):
 \begin{equation}
  p_{M,N} = K_{M,N}^{-1} \prod_{i=1}^M \left( \lambda_i^{N-M} e^{-\lambda_i} \right) \prod_{i<j}(\lambda_i-\lambda_j)^2,
  \label{eq:p-wish}
 \end{equation}
  where $K_{M,N}$ is a normalizing constant.
 We now try to separate the integral in \eqref{eq:prob_sum_first_ev} in two integrals, one over $\lambda_1, \ldots, \lambda_m$, the other over $\lambda_{m+1}, \ldots, \lambda_M$.
 As we have $(\lambda_1, \ldots, \lambda_m) \in \mathcal{C}_m$ in \eqref{eq:prob_sum_first_ev}, $\lambda_m < b / \rho$ and thus
 \begin{equation}
  \begin{split}
      \int_{\lambda_m \leq \lambda_{m+1} \leq \ldots \leq \lambda_M} & p_{M,N}(\lambda_1, \ldots, \lambda_M) \ d\lambda_{m+1} \ldots d\lambda_M
      \\
      & \geq \int_{(\lambda_{m+1}, \ldots, \lambda_M) \in \mathcal{D}} p_{M,N}(\lambda_1, \ldots, \lambda_M) \ d\lambda_{m+1} \ldots d\lambda_M
  \end{split}
  \label{ineq:2intg}
 \end{equation}
 where $\mathcal{D}=\{(\lambda_{m+1}, \ldots, \lambda_M) \in \mathbb{R}_+^{M-m}; \ b/\rho \leq \lambda_{m+1}\leq \ldots \leq \lambda_M \}$.
 This integral can be simplified by noticing that $p_{M,N}(\lambda_1, \ldots, \lambda_M)$ explicit expression \eqref{eq:p-wish} is invariant by permutation of its parameters
 $\lambda_1, \ldots, \lambda_M$,
 in particular by permutation of its parameters
 $\lambda_{m+1}, \ldots, \lambda_M$.
 Therefore, noting
 $\mathcal{S}=\mathrm{Sym}(\{\lambda_{m+1}, \ldots, \lambda_M\})$ the group of permutations over the finite set $\{\lambda_{m+1}, \ldots, \lambda_M\}$,
 we get
 \begin{align}
    \int_{b/\rho}^{+\infty} \bhs\ldots \int_{b/\rho}^{+\infty} & p_{M,N}(\lambda_1, \ldots, \lambda_M) \ d\lambda_{m+1} \ldots d\lambda_M \notag
    \\
    & = \sum_{s \in \mathcal{S}} \int_{s(\lambda_{m+1}, \ldots, \lambda_M) \in \mathcal{D}} p_{M,N}(\lambda_1, \ldots, \lambda_M) \ d\lambda_{m+1} \ldots d\lambda_M \notag
    \\
    & = \mathrm{Card}(\mathcal{S}) \int_{ (\lambda_{m+1}, \ldots, \lambda_M) \in \mathcal{D}} p_{M,N}(\lambda_1, \ldots, \lambda_M) \ d\lambda_{m+1} \ldots d\lambda_M \notag
    \\
    & = (M-m)! \int_{ (\lambda_{m+1}, \ldots, \lambda_M) \in \mathcal{D}} p_{M,N}(\lambda_1, \ldots, \lambda_M) \ d\lambda_{m+1} \ldots d\lambda_M.
    \label{eq:intg-symm}
\end{align}
Using \eqref{ineq:2intg} and \eqref{eq:intg-symm} in \eqref{eq:prob_sum_first_ev}, we obtain
 \[
  \PP \Bigg( \sum_{k=1}^m \rho \lambda_k < b \Bigg)
  \geq \frac{1}{(M-m)!} \int_{\mathcal{C}_m} \int_{b/\rho}^{+\infty} \bhs\ldots \int_{b/\rho}^{+\infty} p_{M,N}(\lambda_1, \ldots, \lambda_M) \ d\lambda_1 \ldots d\lambda_M.
 \]
 We now replace
 $p_{M,N}$
 by its explicit expression \eqref{eq:p-wish} and then try to separate the $m$ first eigenvalues from the others.
 Note that we can drop the constants $(M-m)!$ and $K_{M,N}$ as we only need an asymptotic lower bound.
 \begin{align*}
  \PP \Bigg( \sum_{k=1}^m \rho \lambda_k < b \mhs \Bigg) \mhs
 \ \dot\geq & \int_{\mathcal{C}_m} \int_{b/\rho}^{+\infty} \bhs \ldots \int_{b/\rho}^{+\infty} \prod_{i=1}^M \left( \lambda_i^{N-M} e^{-\lambda_i} \right) \prod_{i<j}(\lambda_i-\lambda_j)^2 \,
 d\lambda_1 \ldots d\lambda_M
 \\
 = & \ \int_{\mathcal{C}_m} \int_{b/\rho}^{+\infty} \bhs \ldots \int_{b/\rho}^{+\infty}
 \Bigg( \prod_{i=1}^m \left( \lambda_i^{N-M} e^{-\lambda_i} \right) \prod_{i<j\leq m}(\lambda_i-\lambda_j)^2 \, \Bigg)
 \\
 & \cdot \Bigg( \prod_{i=m+1}^M \left( \lambda_i^{N-M} e^{-\lambda_i} \right) \prod_{i\leq m<j}(\lambda_i-\lambda_j)^2 \prod_{m<i<j}(\lambda_i-\lambda_j)^2 \,  \Bigg)
 \ d\lambda_1 \ldots d\lambda_M
 \end{align*}
 For $i\leq m<j$, we have that $\lambda_i \leq b/\rho$ and thus $(\lambda_i-\lambda_j)^2 \geq \big(\lambda_j-\frac{b}{\rho}\big)^2$.
 Hence,
 \begin{align}
  \PP \Bigg( \sum_{k=1}^m \rho \lambda_k < b \mhs \Bigg)
  \dot\geq & \  \bigg( \int_{\mathcal{C}_m} \prod_{i=1}^m \left( \lambda_i^{N-M} e^{-\lambda_i} \right) \prod_{i<j\leq m}(\lambda_i-\lambda_j)^2 \   d\lambda_1 \ldots d\lambda_m \bigg)
  \label{eq:2sep-intg}
  \\
  & \cdot \mhs \bigg( \mhs \int_{b/\rho}^{+\infty} \bhs\mhs\mhs \ldots \int_{b/\rho}^{+\infty} 
  \mhs\mhs \prod_{i=m+1}^M  \mhs\mhs\mhs \left( \lambda_i^{N-M}e^{-\lambda_i} \right)
  \mhs \prod_{j=m+1}^M \mhs\mhs\mhs \left(\lambda_j-\frac{b}{\rho}\right)^{\,\mhs\mhs 2m}
  \mhs\mhs \prod_{m<i<j} \mhs\mhs\mhs (\lambda_i-\lambda_j)^2  \  d\lambda_{m+1} \ldots d\lambda_M \mhs\mhs\bigg) \notag
 \end{align}
 We now have two separate integrals.
 We first consider the second one, in which we make the substitution
 $\beta_i= \lambda_i - b/\rho$ for $i=m+1, \ldots, M$.
 \begin{align}
   \int_{b/\rho}^{+\infty} & \bhs \ldots \int_{b/\rho}^{+\infty} 
    \prod_{i=m+1}^M  \mhs\mhs\mhs \left( \lambda_i^{N-M}e^{-\lambda_i} \right)
    \mhs \prod_{j=m+1}^M \mhs\mhs\mhs \left(\lambda_j-\frac{b}{\rho}\right)^{\,\mhs\mhs 2m}
    \mhs\mhs \prod_{m<i<j} \mhs\mhs\mhs (\lambda_i-\lambda_j)^2  \  d\lambda_{m+1} \ldots d\lambda_M
   \notag
   \\
   & =  e^{- {(M-m)b/\rho}} \int_0^{+\infty} \bhs \ldots \int_0^{+\infty}
   \mhs\mhs \prod_{i=m+1}^M \mhs\mhs \left( \mhs \left( {\textstyle \beta_i + \frac{b}{\rho}} \right)^{N-M} e^{- \beta_i} \beta_i ^{2m} \mhs \right)
   \mhs \prod_{m<i<j}(\beta_i-\beta_j)^2 \  d\beta_{m+1} \ldots d\beta_M
   \notag
   \\
   & \geq \frac{1}{2} \int_0^{+\infty} \bhs \ldots \int_0^{+\infty} \prod_{i=m+1}^M \left( \beta_i ^{N-M+2m} e^{- \beta_i} \right) \prod_{m<i<j}(\beta_i-\beta_j)^2 \  d\beta_{m+1} \ldots d\beta_M
   \label{ineq:2nd-intg}
 \end{align}
 for $\rho$ large enough, i.e. such that
 $e^{- (M-m) b/\rho} > 1/2$.
 It is straightforward to see that the integral in \eqref{ineq:2nd-intg} is nonzero, finite, independent from $\rho$ and therefore asymptotically equivalent to $1$ in the sense of \eqref{eq:exp-equ}.
 Hence, we can drop the second integral in \eqref{eq:2sep-intg}, leading to:
 \begin{equation}
    \PP \Bigg( \sum_{k=1}^m \rho \lambda_k < b \mhs \Bigg)
    \dot\geq \  \int_{\mathcal{C}_m} \prod_{i=1}^m \left( \lambda_i^{N-M} e^{-\lambda_i} \right) \prod_{i<j\leq m}(\lambda_i-\lambda_j)^2 \   d\lambda_1 \ldots d\lambda_m.
    \label{ineq:only1-intg}
 \end{equation}
 Making the substitution $\alpha_i= \rho \lambda_i$ for $i=1, \ldots, m$ in \eqref{ineq:only1-intg}
 and 
 noting
 $\mathcal{C}'_m=\{ \alpha_1, \ldots, \alpha_m: \ 0 < \alpha_1 \leq \ldots \leq \alpha_m, \ \sum_{k=1}^m \alpha_k < b \}$
 we then have
 \begin{align}
  \PP \Bigg(\sum_{k=1}^m \rho \lambda_k < b \Bigg)
  & \, \dot\geq \, \bigg( \rho^{-m-m(N-M)-m(m-1)} \int_{\mathcal{C}'_m} \prod_{i=1}^m \left( \alpha_i^{N-M} e^{-\alpha_i/\rho}  \right) \prod_{i<j\leq m}(\alpha_i-\alpha_j)^2  \  d\alpha_1 \ldots  d\alpha_m \bigg) \notag
  \\
  & \geq \rho^{-m(N-M+m)} \int_{\mathcal{C}'_m} \prod_{i=1}^m \left( \alpha_i^{N-M} e^{-\alpha_i} \right) \prod_{i<j\leq m}(\alpha_i-\alpha_j)^2  \   d\alpha_1 \ldots d\alpha_m
  \label{ineq:no-intg}
 \end{align}
 for $\rho \geq 1$, as we have then $e^{- \alpha_i/\rho} \geq e^{- \alpha_i}$ for $i=1, \ldots, m$.
 As $b>0$ it is straightforward to see that the integral in \eqref{ineq:no-intg} is nonzero but also finite and independent from $\rho$; it is therefore asymptotically equivalent to $1$ in the sense of \eqref{eq:exp-equ}, yielding
 \[
  \PP \Bigg( \sum_{k=1}^m \rho \lambda_k < b \Bigg) \dot\geq \ \rho^{-m(N-M+m)},
 \]
 which concludes the proof.

\section{}
\label{apx:prob_first_ev_bounded}
 We prove in this section that $\PP\left( \mathcal{B}_1 \right) \dot\leq \, \rho^{-m(M-N+m)}$, where the set $\mathcal{B}_1$ is defined by
 \[
    \mathcal{B}_1= \{ \lambda_1 \leq \lambda_2 \ldots \leq \lambda_M, \, 0 < \lambda_k \leq b, \, k=1,\ldots,m \},
 \]
 with $b>0$ and $\lambda_1, \ldots, \lambda_M$ the ordered eigenvalues of the Wishart matrix $\H^*\H$.
 We use the same approach as in Appendix \ref{apx:prob_sum_first_ev}.
 For we note $p_{M,N}$ the joint probability density function of the ordered eigenvalues of a $M \times M$ Wishart matrix with scale matrix $\I_M$ and $N$ degrees of freedom,
 the probability $\PP(\mathcal{B}_1)$ can be written as
 \[
  \PP(\mathcal{B}_1)
   = \int_{(\lambda_1, \ldots, \lambda_M) \in \mathcal{B}_1}
   p_{M,N}(\lambda_1, \ldots, \lambda_M) \ d\lambda_1 \ldots d\lambda_M.
 \]
 Similarly to Appendix \ref{apx:prob_sum_first_ev} we try to upper bound $\PP(\mathcal{B}_1)$ by the product of two integrals, one containing the $m$ first eigenvalues and the other the $M-m$ remaining eigenvalues.
 We first replace $p_{M,N}$ by it explicit expression \eqref{eq:p-wish}:
 \begin{align*}
  \PP(\mathcal{B}_1)
  & = K_{M,N}^{-1} \int_{(\lambda_1, \ldots, \lambda_M) \in \mathcal{B}_1}
   \prod_{i=1}^M \lambda_i^{N-M} e^{-\lambda_i} \prod_{i<j}(\lambda_i-\lambda_j)^2 \ d\lambda_1 \ldots d\lambda_M
   \\
  & \doteq \int_{(\lambda_1, \ldots, \lambda_M) \in \mathcal{B}_1}
   \Bigg( \prod_{i=1}^m \left( \lambda_i^{N-M} e^{-\lambda_i} \right) \prod_{i<j\leq m}(\lambda_i-\lambda_j)^2 \Bigg)
   \\
  & \fhs \cdot \Bigg( \prod_{i=m+1}^M \left( \lambda_i^{N-M} e^{-\lambda_i} \right) \prod_{i\leq m<j}(\lambda_i-\lambda_j)^2 \prod_{m<i<j}(\lambda_i-\lambda_j)^2  \Bigg)
  \ d\lambda_1 \ldots d\lambda_M.
 \end{align*}
 Note that we dropped the normalizing constant $K_{M,N}$, as $K_{M,N}^{-1} \doteq 1$.
 For $i \leq m < j$, we have
 $|\lambda_i - \lambda_j| \leq \lambda_j$
 and thus
 $\prod_{i\leq m<j}(\lambda_i-\lambda_j)^2 \leq \prod_{j=m+1}^M \lambda_j^{2m}$, yielding
 \begin{align*}
  \PP(\mathcal{B}_1)
  & \,\dot\leq\,
    \int_0^{b/\rho} \int_{\lambda_1}^{b/\rho} \bhs \ldots \int_{\lambda_{m-1}}^{b/\rho} \int_{\lambda_m}^{+\infty} \bhs \ldots \int_{\lambda_{M-1}}^{+\infty}
    \Bigg( \prod_{i=1}^m \left( \lambda_i^{N-M} e^{-\lambda_i} \right) \prod_{i<j\leq m}(\lambda_i-\lambda_j)^2 \Bigg)
    \\
  & \fhs \cdot  \Bigg( \prod_{i=m+1}^M \left( \lambda_i^{N+2m-M} e^{-\lambda_i} \right) \prod_{m<i<j}(\lambda_i-\lambda_j)^2  \Bigg)
    \ d\lambda_1 \ldots d\lambda_M
 \end{align*}
 In order to obtain two separate integrals we discard the $\lambda_m$ in the integral bound simply by noticing that $\lambda_m>0$, therefore
 \begin{align*}
  \PP(\mathcal{B}_1)
  & \,\dot\leq\,
    \Bigg( \int_0^{b/\rho} \int_{\lambda_1}^{b/\rho} \bhs \ldots \int_{\lambda_{m-1}}^{b/\rho}
    \prod_{i=1}^m \left( \lambda_i^{N-M} e^{-\lambda_i} \right) \prod_{i<j\leq m}(\lambda_i-\lambda_j)^2
    \ d\lambda_1 \ldots d\lambda_m \Bigg)
    \\
  & \fhs \cdot \Bigg( \int_0^{+\infty} \int_{\lambda_{m+1}}^{+\infty} \bhs \ldots \int_{\lambda_{M-1}}^{+\infty}
    \prod_{i=m+1}^M \left( \lambda_i^{N+2m-M} e^{-\lambda_i} \right) \prod_{m<i<j}(\lambda_i-\lambda_j)^2
    \ d\lambda_{m+1} \ldots d\lambda_M \Bigg)
 \end{align*}
 As the second integral (in $\lambda_{m+1}$, \ldots, $\lambda_M$) is nonzero, finite and independent of $\rho$ it is asymptotically equivalent to $1$ in the sense of \eqref{eq:exp-equ}.
 Hence,
 \begin{equation}
  \PP(\mathcal{B}_1)
  \,\dot\leq\,
    \int_0^{b/\rho} \int_{\lambda_1}^{b/\rho} \bhs \ldots \int_{\lambda_{m-1}}^{b/\rho}
    \prod_{i=1}^m \left( \lambda_i^{N-M} e^{-\lambda_i} \right) \prod_{i<j\leq m}(\lambda_i-\lambda_j)^2
    \ d\lambda_1 \ldots d\lambda_m.
    \label{ineq:PB1-1intg}
 \end{equation}
 We now make the substitutions $\alpha_i=\rho\lambda_i$ for $i=1, \ldots, m$ inside the remaining integral.
 \begin{align}
  & \int_0^{b/\rho} \int_{\lambda_1}^{b/\rho} \bhs \ldots \int_{\lambda_{m-1}}^{b/\rho}
    \prod_{i=1}^m \left( \lambda_i^{N-M} e^{-\lambda_i} \right) \prod_{i<j\leq m}(\lambda_i-\lambda_j)^2
    \ d\lambda_1 \ldots d\lambda_m \notag
  \\
  & \hspace{15pt} = \ \rho^{-m(N-M+m)}
    \int_0^{b} \int_{\alpha_1}^{b} \bhs \ldots \int_{\alpha_{m-1}}^{b}
    \prod_{i=1}^m \left( \alpha_i^{N-M} e^{-\alpha_i/\rho} \right) \prod_{i<j\leq m}(\alpha_i-\alpha_j)^2
    \ d\alpha_1 \ldots d\alpha_m \notag
  \\
  & \hspace{15pt} \leq  \rho^{-m(N-M+m)} 
    \mhs \int_0^{b} \int_{\alpha_1}^{b} \bhs \ldots \int_{\alpha_{m-1}}^{b}
    \prod_{i=1}^m  \alpha_i^{N-M}  \mhs \prod_{i<j\leq m} \mhs\mhs (\alpha_i-\alpha_j)^2 
    \ d\alpha_1 \ldots d\alpha_m , \label{ineq:rem_intg}
 \end{align}
 as $e^{-\alpha_i/\rho} \leq 1$.
 The remaining integral in \eqref{ineq:rem_intg} is nonzero ($b > 0$), finite and does not depend on $\rho$; therefore, \eqref{ineq:rem_intg} is asymptotically equivalent to $\rho^{-m(N-M+m)}$ in the sense of \eqref{eq:exp-equ}.
 Coming back to \eqref{ineq:PB1-1intg} we obtain
 \[
  \PP(\mathcal{B}_1) \ \dot\leq \ \rho^{-m(N-M+m)}.
 \]

\section{}
\label{apx:ang_par}
 In this appendix, we review the results of \cite{dita2003factorization, lundberg2004haar} for the reader's convenience.
 
 It has been shown in \cite{dita2003factorization} that any $n \times n$ unitary matrix $A_n$ can be written as
 \begin{equation}
  A_n = d_n {\mathcal{O}}_n \begin{bmatrix} 1 & 0 \\ 0 & A_{n-1} \end{bmatrix},
  \label{eq:unitary_dec}
 \end{equation}
 with
 $A_{n-1}$ a $(n-1) \times (n-1)$ unitary matrix,
 $d_n$ a diagonal phases matrix, that is
 $d_n = \mathrm{diag}(e^{i\varphi_1}, \ldots, e^{i\varphi_n})$ with $\varphi_1, \ldots, \varphi_n \in [0, 2\pi]$,
 and
 $\mathcal{O}_n$ an orthogonal matrix (the angles matrix).
 Matrix $\mathcal{O}_n$ can be written in terms of parameters $\theta_1, \ldots, \theta_n \in [0, \frac{\pi}{2}]$ thanks to the following decomposition:
 $\mathcal{O}_n = J_{n-1,n} J_{n-2,n-1} \ldots J_{1,2}$,
 where
 \[
    J_{i,i+1}= \begin{bmatrix} \I_{i-1} & 0 & 0 & 0 \\ 0 & \cos\theta_i & -\sin\theta_i & 0 \\ 0 & \cos\theta_i & -\sin\theta_i & 0 \\ 0 & 0 & 0 & \I_{n-i-1} \end{bmatrix}.
 \]

 Let $\U_M$ be a $M \times M$ unitary Haar distributed matrix.
 Then, using decomposition \eqref{eq:unitary_dec},
 \[
  \U_M= \D_M(\bs\varphi_1) \V_M(\bs\theta_1) \begin{bmatrix} 1 & 0 \\ 0 & \U_{M-1} \end{bmatrix},
 \]
 with
 $\bs\varphi_1=(\varphi_{1,1},\ldots,\varphi_{1,M}) \in [0,2\pi]^M$,
 $\bs\theta_1=(\theta_{1,1},\ldots,\theta_{1,M-1}) \in [0,\frac{\pi}{2}]^{M-1}$,
 $\D_M(\bs\varphi_1)$ the diagonal matrix defined by $\D_M(\bs\varphi_1)=\mathrm{diag}(e^{i\varphi_{1,1}},\ldots,e^{i\varphi_{1,M}})$,
 $\V_M(\bs\theta_1)$ the orthogonal matrix defined by $\V_M(\bs\theta_1)=J_{M-1,M} J_{M-2,M-1} \ldots J_{1,2}$
 and
 $\U_M$ a $M-1 \times M-1$ unitary matrix.
 Matrix $\U_{M-1}$ can naturally be similarly factorized.
 
 Similarly to \cite{lundberg2004haar}, we can show that,
 in order $\U_M$ to be a Haar matrix it is sufficient
 that $(\varphi_{1,i})_{i=1,\ldots,M}$ are i.i.d. random variables uniformly distributed over interval $[0, 2\pi[$,
 that $\theta_{1,1},\ldots,\theta_{1,M-1}$ are independent with densities respectively equal to $(\sin \theta_1)^{M-2}, (\sin \theta_2)^{M-3}, \ldots, (\sin \theta_{M-2}), 1$ and independent from $\bs\varphi_1$
 and that $\U_{M-1}$ is Haar distributed and independent from $\bs\varphi_1$ and $\bs\theta_1$.
 The proof consists in first showing, by a simple variable change, that if the $(\varphi_{1,i})_{i=1,\ldots,M}$ and the $\theta_{1,1},\ldots,\theta_{1,M-1}$ follow the mentioned distributions then $\D_M(\bs\varphi_1) \V_M(\bs\theta_1)$ is uniformly distributed over the unity sphere of $\mathbb{C}^M$.
 The proof is then completed by showing that if $\U_{M-1}$ is a Haar matrix independent from $\bs\varphi_1$ and $\theta_1$ then $\U_M$ is Haar distributed.
 
 Finally one can parameterize a Haar matrix $\U_M$ by $\bs\varphi_1$, $\theta_1$ and $\U_{M-1}$.
 Repeating the same parametrization for $\U_{M-1}$ we obtain that $\U_M$ can be parameterized by the $M^2$ following independent variables
 \begin{align*}
  &(\varphi_{1,1}, \ldots, \varphi_{1,M}), (\theta_{1,1}, \ldots, \theta_{1,M-1}), (\varphi_{2,1}, \ldots, \varphi_{2,M-1}), (\theta_{2,1}, \ldots, \theta_{2,M-2}), \ldots,
  \\
  &(\varphi_{M-2,1}, \varphi_{M-2,2}), \theta_{M-2,1}, \varphi_{M-1,1},
 \end{align*}
 whose probability laws are almost surely positive.

\bibliographystyle{IEEEtran}
\bibliography{IEEEabrv,bibMarne}

\begin{thebibliography}{10}
\providecommand{\url}[1]{#1}
\csname url@samestyle\endcsname
\providecommand{\newblock}{\relax}
\providecommand{\bibinfo}[2]{#2}
\providecommand{\BIBentrySTDinterwordspacing}{\spaceskip=0pt\relax}
\providecommand{\BIBentryALTinterwordstretchfactor}{4}
\providecommand{\BIBentryALTinterwordspacing}{\spaceskip=\fontdimen2\font plus
\BIBentryALTinterwordstretchfactor\fontdimen3\font minus
  \fontdimen4\font\relax}
\providecommand{\BIBforeignlanguage}[2]{{%
\expandafter\ifx\csname l@#1\endcsname\relax
\typeout{** WARNING: IEEEtran.bst: No hyphenation pattern has been}%
\typeout{** loaded for the language `#1'. Using the pattern for}%
\typeout{** the default language instead.}%
\else
\language=\csname l@#1\endcsname
\fi
#2}}
\providecommand{\BIBdecl}{\relax}
\BIBdecl

\bibitem{zheng2003diversity}
L.~Zheng and D.~Tse, ``{Diversity and multiplexing: A fundamental tradeoff in
  multiple-antenna channels},'' \emph{{IEEE} Trans. Inform. Theory}, vol.~49,
  no.~5, pp. 1073--1096, 2003.

\bibitem{kumar2009asymptotic}
K.~Kumar, G.~Caire, and A.~Moustakas, ``Asymptotic performance of linear
  receivers in mimo fading channels,'' \emph{{IEEE} Trans. Inform. Theory},
  vol.~55, no.~10, pp. 4398--4418, oct. 2009.

\bibitem{hedayat2005linear}
A.~Hedayat, A.~Nosratinia, and N.~Al-Dhahir, ``Linear equalizers for flat
  rayleigh mimo channels,'' in \emph{Proc. ICASSP Conference}, vol.~3, march
  2005, pp. 445--448.

\bibitem{hedayat2004outage}
------, ``Outage probability and diversity order of linear equalizers in
  frequency-selective fading channels,'' in \emph{Proc. Asilomar Conference},
  vol.~2, nov. 2004, pp. 2032--2036.

\bibitem{tajer2007diversity}
A.~Tajer and A.~Nosratinia, ``Diversity order of mmse single-carrier frequency
  domain linear equalization,'' in \emph{Proc. Globecom Conference}, nov. 2007,
  pp. 1524--1528.

\bibitem{mehana2010diversity}
A.~Mehana and A.~Nosratinia, ``{Diversity of MMSE MIMO receivers},'' in
  \emph{Information Theory Proceedings (ISIT), 2010 IEEE International
  Symposium on}, june 2010, pp. 2163--2167.

\bibitem{mehana2011diversity}
\BIBentryALTinterwordspacing
------, ``Diversity of mmse mimo receivers,'' 2011. [Online]. Available:
  \url{http://arxiv.org/abs/1102.1462}
\BIBentrySTDinterwordspacing

\bibitem{jiang2011performance}
Y.~Jiang, M.~Varanasi, and J.~Li, ``{Performance analysis of ZF and MMSE
  equalizers for MIMO systems: an in-depth study of the high SNR regime},''
  \emph{{IEEE} Trans. Inform. Theory}, vol.~57, no.~4, pp. 2008--2026, apr.
  2011.

\bibitem{dita2003factorization}
P.~Di\c{t}\u{a}, ``Factorization of unitary matrices,'' \emph{J. Phys. A: Math.
  Gen.}, vol.~36, pp. 2781--2789, 2003.

\bibitem{lundberg2004haar}
M.~Lundberg and L.~Svensson, ``{The Haar measure and the generation of random
  unitary matrices},'' in \emph{Proc. IEEE Sensor Array and Multichannel Signal
  Processing Workshop}, 2004, pp. 114--118.

\end{thebibliography}

\end{document}